\documentclass[11pt]{article}
\usepackage{epsfig}
\newcommand{\sss}{\vspace{.2in}}

\def\nonu{\nonumber}
\def\br{\begin{eqnarray}}
\def\er{\end{eqnarray}}
\def\be{\begin{equation}}
\def\ee{\end{equation}}
\def\lb{\lbrack}
\def\rb{\rbrack}

\def\lcurl{\left\{}
\def\rcurl{\right\}}
\def\>{\rangle}              
\def\<{\langle}              
\def\({\left(}
\def\){\right)}
\def\v{\vert}                     

\def\ra{\rightarrow}

\def\f{\frac}

\def\eq{\!\!\!\! &=& \!\!\!\! }
\def\Ad{A^{\dagger}}
\def\jp{J_{+}}
\def\jm{J_{-}}
                                                          \def\j3{J_3}
\def\id{i\partial_\phi}
\def\hm{h_{min}}

\def\a{\alpha}

\def\p{\phi}

\begin{document}
~\hfill{\footnotesize UICHEP-TH/98-7,~~\today}
\sss
\begin{center}
{\Large {\Large \bf Algebraic Shape Invariant Models}}
\end{center}
\vspace{.5in}
\begin{center}
{\large{\bf
   \mbox{S. Chaturvedi}$^{a,}$\footnote{scsp@uohyd.ernet.in},
   \mbox{R. Dutt}$^{b,}$\footnote{rdutt@vbharat.ernet.in},
   \mbox{A. Gangopadhyaya}$^{c,}$\footnote{agangop@luc.edu, asim@uic.edu},
   \mbox{P. Panigrahi}$^{a,}$\footnote{panisp@uohyd.ernet.in},
   \mbox{C. Rasinariu}$^{d,}$\footnote{costel@uic.edu} and
   \mbox{U. Sukhatme}$^{d,}$\footnote{sukhatme@uic.edu}
 }}
\end{center}
\vspace{.6in}
\noindent
a) \hspace*{.2in}
Department of Physics, University of Hyderabad, Hyderabad, India;\\
b) \hspace*{.2in}
Department of Physics, Visva Bharati University, Santiniketan, India;
\\
c) \hspace*{.2in}
Department of Physics, Loyola University Chicago, Chicago, USA; \\
d) \hspace*{.2in}
Department of Physics, University of Illinois at Chicago, Chicago,
USA. \\

\begin{abstract}

Motivated by the shape invariance condition in supersymmetric quantum
mechanics, we develop an algebraic framework for shape invariant
Hamiltonians with a general change of parameters. This approach involves
nonlinear generalizations of Lie algebras. Our work extends previous results
showing the equivalence of shape invariant potentials involving translational
change of parameters with standard $SO(2,1)$ potential algebra for Natanzon
type potentials.
\end{abstract}

\newpage
In recent years, there has been considerable interest in studying exactly
solvable quantum mechanical problems using algebraic
approaches[1-9]. In this respect, supersymmetric quantum mechanics
(SUSYQM) has been found to be an elegant and useful  prescription for 
obtaining closed analytic expressions both for the energy eigenvalues and 
eigenfunctions for a large class of one dimensional (or spherically symmetric 
three dimensional) problems \cite{Cooper}. The main ingredients in SUSYQM are the 
supersymmetric partner Hamiltonians $H_{-} \equiv \Ad A$ and 
$H_{+} \equiv A \Ad$. The $A$ and $\Ad$ operators used in this 
factorization are expressed in terms of the real superpotential $W$ 
as follows:
\be
\label{A}
   A(x,a) = \frac{d}{dx} + W(x,a) \quad;\quad \Ad (x,a) =
-\frac{d}{dx} + W(x,a)~.
\ee
Here, $a$ is a parameter (or a set of parameters), which will play
an important role in the approach of this paper. An
interesting feature of SUSYQM is that
for a shape invariant system [1-3], 
i.e. a system satisfying
\be
\label{si}
   H_{+}(x,a_0) = H_{-}(x,a_1) + R(a_0)~;
   \quad a_1 = f(a_0)~~,
\ee
the entire spectrum can be determined algebraically without ever
referring to underlying differential equations.

Recently, it has been shown 
that shape invariant potentials in which the change of parameters
\mbox{$a_1 = f(a_0)$} is of translational type $a_1 = a_0 + k,~ k =
{\rm constant}$,
possess a $SO(2,1)$ potential algebra and are solvable by group
theoretical
techniques[4-7]. Thus, for
such potentials, there appear to be two seemingly
independent algebraic
methods, SUSYQM and potential algebras, for obtaining the
complete spectra. One may naturally ask the question
whether there is any connection between the two methods. Indeed,
the equivalence between the two algebraic approaches has been
established \cite{Gangopadhyaya_proc}. More specifically, for all shape
invariant potentials of the Natanzon type \cite{Natanzon}, the underlying potential algebra
is $SO(2,1)$. Other similar approaches are discussed in refs.
\cite{Alhassid,Cordero}.

The purpose of this letter is to examine whether such an equivalence
is also valid for a larger class of shape invariant potentials.
We consider a much more general class of change of parameters,
which contains both translations as well as scalings. In particular, for shape
invariant potentials generated by a pure scaling
ansatz \cite{Barclay}
$a_1 =q a_0, q = {\rm constant}~ (0<q<1)$, we find that the
associated potential algebra is a nonlinear deformation of $SO(2,1)$.  In fact, 
coordinate representations of these nonlinear algebras are an interesting area 
of research in their own right from a mathematical point of view 
\cite{Beckers}. We also show that the
energy spectrum resulting from this potential algebra agrees with
the results of SUSYQM.

To begin the construction of the operator algebra, let us
express eq. (\ref{si}) in terms of $A$
and $\Ad$ :
\be
\label{asi}
   A(x,a_0) \Ad (x,a_0) -  \Ad(x,a_1) A (x,a_1) =  R(a_0)~.
\ee
This relation resembles a commutator structure. To obtain a closed
$SO(2,1)$ type Lie algebra, we introduce an auxiliary variable $\phi$ and
define
\br
\label{asim}
    \jp \eq Q(\phi) \Ad(x, \chi(\id)) ~, \nonumber \\
    \jm \eq  A(x, \chi(\id)) \: Q^{\star}(\phi)~,
\er
where $Q(\phi)$ is a function to be determined and $\chi$ is an arbitrary,
real function. The operator
$A(x, \chi(\id))$ is given by eq. ({\ref{A}) with the substitution
\be
\label{chi}
a \ra \chi(\id)~.
\ee
From eq. (\ref{asim}), one
obtains
\be
\label{q}
[\jp,\jm] =  Q(\phi) \Ad(x, \chi(\id))A(x, \chi(\id)) \:Q^{\star}(\phi)
          - A(x, \chi(\id)) \: Q^{\star}(\phi)  Q(\phi) \Ad(x, \chi(\id))~.
\ee
Choosing $Q(\phi)=e^{ip\phi}$, where $p$ is an arbitrary real
constant,
eq. (\ref{q}) can be easily cast into the following form
\be
\label{j+/-}
  [\jp,\jm] = -\lcurl A(x, \chi(\id))\Ad(x, \chi(\id))
          - \Ad(x, \chi(\id + p))A(x, \chi(\id + p)) \rcurl~.
\ee
In analogy with eq. (\ref{asi}), we impose the algebraic shape
invariance condition
\be
  A(x, \chi(\id))\Ad(x, \chi(\id))
  - \Ad(x, \chi(\id + p))A(x, \chi(\id + p)) = R(\chi(\id))~.
\ee
Note that the quantity $J_+J_-$ corresponds to the Hamiltonian
\be \label{nn}
H_-(x,\id+p)= \Ad(x, \chi(\id + p))A(x, \chi(\id + p))~.
\ee
Identifying
\be
\label{j3}
  \j3 = -\frac{\id}{p} ~,
\ee
we obtain a deformed $SO(2,1)$ Lie algebra whose generators
$\jp, \jm$ and $\j3$ satisfy
\br
\label{lie}
   \lb \j3, J_{\pm} \rb \eq \pm \;J_{\pm}~, \nonumber \\
   \lb \jp,\jm \rb \eq \xi(\j3)~.
\er
Here $\xi(\j3) \equiv -R(\chi(\id))$.
The algebraic shape invariance condition (\ref{asi}) is the main
tool of our analysis. If we identify
\be
\label{a01}
  a_0 \ra \chi(\id) \quad ; \quad a_1 \ra \chi(\id+p)~,
\ee
then depending on the choice of the $\chi$ function in eq. (\ref{a01})
we have:
\begin{enumerate}
    \item translational models: $a_1=a_0+p$ for $\chi(z)=z$~,
    \item scaling models: $a_1= e^p a_0 \equiv q a_0$ for
$\chi(z)=e^z$~.
\end{enumerate}

One may note that in the translational reparametrization ($\chi(z)=z$)
the commutation relations (\ref{lie}) describe a standard,
undeformed $SO(2,1)$ Lie algebra with $\xi(J_3)=-2J_3$,
as found in ref. \cite{Gangopadhyaya_proc}. Other changes of parameters
follow from more complicated choices for $\chi(z)$. For example, if one
takes $\chi(z)=e^{e^z}$, one gets the change of parameters $a_1=a_0^2$. 
Thus, with judicious choices for $\chi(z)$, various other 
reparametrizations can be emulated. 

In this letter, we will analyze the potential algebra of scaling type reparametrizations. Different models with such multiplicative reparametrizations are characterized by the form of the remainder function $R(a_0)$. As a first example of scaling, we choose $R(a_0)=R_1a_0$. This choice 
generates self-similar potentials studied in refs. \cite{Barclay,Spiridinov}.  
In this case, eqs. (\ref{lie}) become:
\be
\label{scaling}
   \lb \j3, J_{\pm} \rb = \pm \;J_{\pm} \quad ; \quad
   \lb \jp,\jm \rb = \xi(\j3) \equiv  -R_1~\exp(-p\j3) ~.
\ee
Thus, we see that the resulting potential algebra is no longer given by a
$SO(2,1)$ Lie algebra, but is deformed. To find the energy
spectrum of the Hamiltonian
$H_{-}$ of eq. (\ref{nn}), we first construct 
the unitary representations of this
deformed Lie algebra
\cite{Rocek}. We define, up to an additive constant, a function
$g(\j3)$ such that
\be
\label{g-def}
   \xi(\j3) = g(\j3)-g(\j3 - 1) ~.
\ee
The Casimir of this algebra is then given $C_2=\jm \jp + g(\j3)$.
In a basis in which $\j3$ and $C_2$ are diagonal, $\jp$
and $\jm$ play the role of raising and lowering operators, respectively.
Operating on an eigenstate $\v h \rangle $ of $J_3$, we have
\br
\label{state}
  \j3 \v h\> \eq h \v h\>~, \nonumber \\
  \jm \v h\> \eq a(h) \; \v h-1\>~, \nonumber \\
  \jp \v h\> \eq a^{\star}(h+1)\; \v h+1\> ~.
\er
Using eqs. (\ref{lie}) and (\ref{state}) we obtain
\be
\label{a}
\v a(h) \v^2 - \v a(h+1) \v^2 = g(h) - g(h-1) ~.
\ee

The profile of $g(h)$ determines the dimension of the unitary representation.
To illustrate how this mechanism works, let us consider the two cases
presented in fig. \ref{g-plot}.
\begin{figure}[ht]
\centering
    \epsfig{file=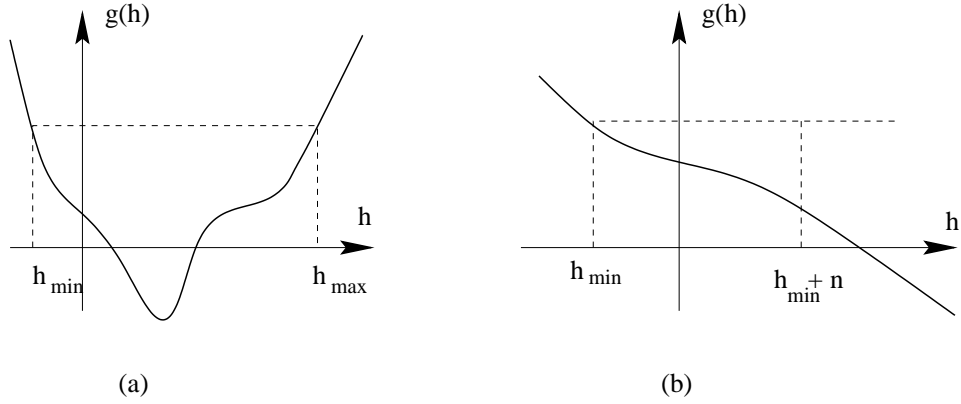, width=5in}
    \caption{Generic behaviors of $g(h)$.}
    \label{g-plot}
\end{figure}
One obtains finite dimensional representations fig. 1a, by starting
from a point on the \mbox{$g(h)$ vs. $h$} graph corresponding
to $h=\hm$, and moving in integer steps parallel to the $h$-axis till the
point corresponding to $h=h_{max}$. Only those discrete 
values of $\hm$ are allowed for which at the end points
$a(\hm)=a(h_{max}+1)=0$, and we get a finite representation.
This is the case of $SU(2)$ for example, where $g(h)$ is
given by the parabola $h(h+1)$. For a monotonically decreasing function $g(h)$,
fig. 1b, there exists only one end point at $h=\hm$. Starting from $\hm$ the
value of $h$ can be increased in integer steps till infinity. 
In this case we have an infinite dimensional representation. 
As in the finite case, $\hm$ labels the representation. 
However, the difference is that here $\hm$ takes continuous values.
Similar arguments apply for a monotonically increasing function $g(h)$.
A more detailed discussion of these techniques is given in ref. \cite{Rocek}.

In our case, from eqs. (\ref{scaling}) and (\ref{g-def}) one gets
\be
\label{g}
    g(h) = \frac{R_1}{e^p - 1} e^{-p h} = -\frac{R_1}{1-q} q^{- h} 
    \quad;\quad q=e^p~,
\ee
represented graphically in fig. 2. Note that for scaling problems \cite{Barclay}, 
one requires $0<q<1$, which leads to $p<0$. 
From the monotonically decreasing profile of the function $g(h)$, it follows 
that the unitary representations of this algebra are infinite dimensional.
If we label the lowest weight state of the operator $\j3$ by $\hm$,
then $a(\hm)=0$. Without loss of generality we can choose the
coefficients $a(h)$ to be real. Then one obtains from (\ref{a}) for an
arbitrary \mbox{$h = \hm + n,~ n=0,1,2,{\ldots} $} 
\be
\label{aj}
  a^2(h) = g(h - n - 1) - g(h - 1) =
R_1~ \frac{q^n-1}{q-1} \: q^{1-h}
\ee
where $q = e^p$.
The spectrum of the Hamiltonian $H_-(x,a_0)$ is given by
\be
\label{spectrum}
H_-\v h \> = a^2(h) \v h \> = R_1~ \frac{q^n-1}{q-1} \: q^{1 - h}\v h \>~.
\ee
Therefore, the eigenenergy is:
\be
\label{energy}
E_n(h) = ~\a(h) \frac{q^n-1}{q-1} \quad ;\quad \a(h) \equiv R_1~q^{1-h} ~. 
\ee
It is interesting to note that the eigenvalue $h$ of the operator $J_3$ plays the role 
of a parameter for the theory. To compare the spectrum of this 
Hamiltonian obtained in eq. (\ref{spectrum}) with those obtained from 
SUSYQM \cite{Barclay}, we go to the
$x$-representation. Here  $\v h \> \propto e^{ip h \phi}\psi_{\hm,n}(x)$ and
hence, the Schr\"odinger equation for the Hamiltonian $H_-$ reads
$$
\lcurl -\frac{d^2}{dx^2} + W^2(x,e^{\id +p}) - 
      W'(x,e^{\id +p}) -E \rcurl \: e^{ip\p h}\psi_{\hm,n}(x) = 0 ~,
$$
or
\be
\lcurl -\frac{d^2}{dx^2} + W^2(x,\a) - W'(x,\a) -E \rcurl
\psi_{\hm,n}(x) = 0 ~,
\ee
which is exactly the Schr\"odinger equation for the scaling model in SUSYQM. 
Eigenenergies of this Hamiltonian \cite{Barclay} are indeed given by 
eq. (\ref{energy}).
Hence, the energy $E$ is given by eq. (\ref{energy}) for $h$ = constant. 
In fig. 2, we depict the elegant correspondence that exists between 
potential algebra and supersymmetric quantum mechanics for shape invariant
potentials.
\begin{figure}[ht]
\centering
    \epsfig{file=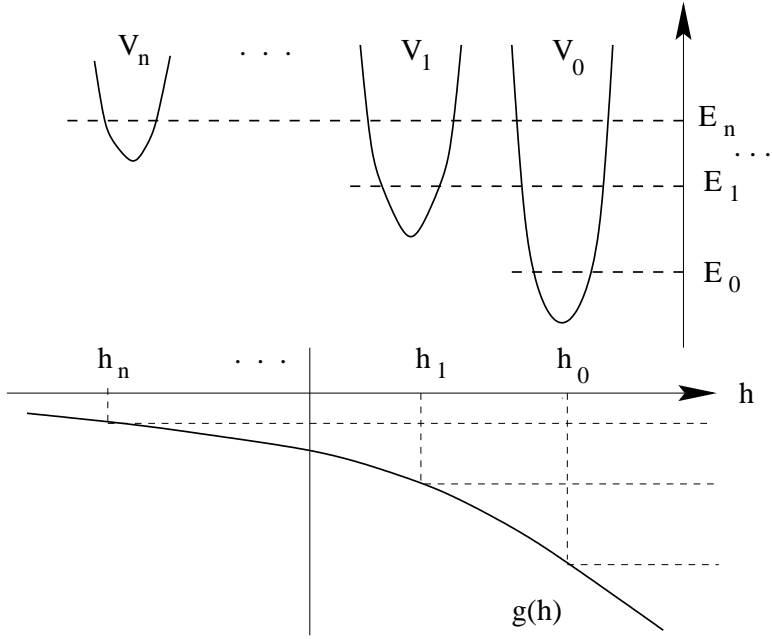, width=4in}
    \caption{Schematic diagram showing the correspondence between potential algebra and SUSYQM with scaling type reparametrization.}
    \label{potentials}
\end{figure}
Let us fix $h=\hm + n$ to a given value, say $h_0$.
For $n=0$ we have a representation of our deformed SO(2,1) algebra labeled by
$h_0 \equiv \hm$. We denote the corresponding potential from the Schr\"odinger
equation of this Hamiltonian by $V_0$. From eq. (\ref{energy}) the
energy $E_0(h_0) = 0$ as we should expect for the ground state in SUSYQM.
For $n=1$, we obtain the first excited state for $V_0$ at $E_1(h_0)=\a(h_0)$.
Note that we can write $h_0=(\hm - 1) + 1 \equiv h_1 + 1$. Construct
now a new representation labeled by $h_1 \equiv h_0 - 1$. The corresponding
potential $V_1$ has its ground state energy, at the same level with
the first excited state of $V_0$. But this is the well known property of the
supersymmetric partner potentials. One can now iterate this procedure.
After $n$ steps, we build a representation labeled by $h_n$ where
$h_n = h_0 - n$, and the ground state energy of the corresponding potential
$V_n$, will be exactly the $n$--excited state energy of $V_0$. 

For a more general case, we assume $R(a_0)=\sum_{j=1}^\infty R_j a_0^j$. In 
this case 
\be
g(h)=\sum_{j=1}^\infty \f{R_j}{1-e^{jp}}e^{-jph}~~,
\ee
and therefore one gets
\br
a^2(h) \eq g(h-n-1)-g(h-1) \nonu \\
\eq \sum_{j=1}^\infty \a_j (h) \f{1-q^{jn}}{1-q^j}~~,
\er
where $\a_j (h) = R_j e^{-j(h-1)}$. 
This result agrees with those obtained in ref. \cite{Barclay}.

In conclusion, in this paper we have developed an algebraic approach involving 
nonlinear generalizations of Lie algebras for treating general shape 
invariant Hamiltonians.


A.G. acknowledges a research leave and a grant from Loyola University 
Chicago which made his involvement in this work possible. R.D. 
would also like to thank the Physics Department of the University of
Illinois at Chicago for warm hospitality. Partial financial support from
the U.S. Department of Energy is gratefully acknowledged.
\newpage

\end{document}